\documentclass[aps,prc,twocolumn,nofootinbib,amsmath,showpacs,superscriptaddress,floatfix]{revtex4}
\usepackage{graphicx,epsfig,longtable}
\usepackage{mathptmx}
\usepackage[light,first]{draftcopy} 
\usepackage{epstopdf}  
\usepackage[pdftex]{hyperref}
\usepackage{wasysym}

\newcommand{\beqy}{\begin{eqnarray}}
\newcommand{\eeqy}{\end{eqnarray}}
\newcommand{\bmlet}{\begin{subequations}}
\newcommand{\emlet}{\end{subequations}}
\newcounter{saveeqn}

\def\gsimeq{\,\,\raise0.14em\hbox{$>$}\kern-0.76em\lower0.28em\hbox  
{$\sim$}\,\,}  
\def\lsimeq{\,\,\raise0.14em\hbox{$<$}\kern-0.76em\lower0.28em\hbox  
{$\sim$}\,\,}  

\begin{document}

\title{Shell-gap reduced level densities in $^{89,90}$Y}

\author{M.~Guttormsen}
\email{magne.guttormsen@fys.uio.no}
\affiliation{Department of Physics, University of Oslo, N-0316 Oslo, Norway}
\author{A.C.~Larsen}
\affiliation{Department of Physics, University of Oslo, N-0316 Oslo, Norway}
\author{F.L.~Bello Garrote}
\affiliation{Department of Physics, University of Oslo, N-0316 Oslo, Norway}
\author{Y.~Byun}
\affiliation{Department of Physics and Astronomy, Ohio University, Athens, Ohio 45701, USA}
\author{T.K.~Eriksen}
\affiliation{Department of Physics, University of Oslo, N-0316 Oslo, Norway}
\author{F.~Giacoppo}
\affiliation{Department of Physics, University of Oslo, N-0316 Oslo, Norway}
\author{A.~G{\"o}rgen}
\affiliation{Department of Physics, University of Oslo, N-0316 Oslo, Norway}
\author{T.W.~Hagen}
\affiliation{Department of Physics, University of Oslo, N-0316 Oslo, Norway}
\author{M.~Klintefjord}
\affiliation{Department of Physics, University of Oslo, N-0316 Oslo, Norway}
\author{H.T.~Nyhus}
\affiliation{Department of Physics, University of Oslo, N-0316 Oslo, Norway}
\author{T.~Renstr{\o}m}
\affiliation{Department of Physics, University of Oslo, N-0316 Oslo, Norway}
\author{S.J.~Rose}
\affiliation{Department of Physics, University of Oslo, N-0316 Oslo, Norway}
\author{E.~Sahin}
\affiliation{Department of Physics, University of Oslo, N-0316 Oslo, Norway}
\author{S.~Siem}
\affiliation{Department of Physics, University of Oslo, N-0316 Oslo, Norway}
\author{T.~Tornyi}
\affiliation{Department of Physics, University of Oslo, N-0316 Oslo, Norway}
\author{G.M.~Tveten}
\affiliation{Department of Physics, University of Oslo, N-0316 Oslo, Norway}
\author{A.~Voinov}
\affiliation{Department of Physics and Astronomy, Ohio University, Athens, Ohio 45701, USA}

\date{\today}

\begin{abstract}
Particle-$\gamma$ coincidences from the $^{89}$Y($p,p' \gamma$)$^{89}$Y
and $^{89}$Y($d,p \gamma$)$^{90}$Y reactions were utilized to obtain $\gamma$-ray
spectra as function of excitation energy. The Oslo method was used
to extract the level density from the particle-$\gamma$ coincidence matrices.
The impact of the $N=50$ shell closure on the level densities is discussed
within the framework of a combinatorial quasi-particle model.
\end{abstract}


\maketitle

\section{Introduction}
\label{sec:int}
Experimental level densities in the quasi-continuum of atomic nuclei represent an important test ground for nuclear structure models.
They contain information on the average distance between single-particle energy levels, the size of shell gaps and residual interactions
like the pairing force between nucleons in time-reversed orbitals. Level densities also play an essential role in the calculation of
reaction cross sections for various applications such as  astrophysical nucleosynthesis, nuclear energy production and transmutation
of nuclear waste.

The total level density $\rho(E) = \sum_{I{\pi}}\rho(E,I,\pi)$ at an excitation energy $E$ depends both on the spin ($I$) 
and parity ($\pi$) distributions. The density of single-particle orbitals becomes strongly reduced at nuclear shell gaps. 
Since the total level density directly depends on available orbitals around the Fermi surface, dramatic effects are expected 
to occur in the vicinity of closed shells, in particular for the spin and parity distributions.
It was recently pointed out~\cite{luciano2014} that pairing as well as shell gaps give a constant-temperature 
level density and that this behavior is a direct evidence for a first-order phase transition.
In this work we study how the level densities are affected by the low single-particle level density at shell gaps, 
both as function of neutron number and excitation energy.
    
At low excitation energy, the nuclear level density can be reliably determined from the counting of low-lying discrete 
known levels~\cite{NNDC}. There is also valuable level-density information from neutron resonance energy spacings 
at the neutron separation energies~\cite{RIPL3}. However, the level density in between these excitation regions 
is for many nuclei a {\em terra incognita}.

The Oslo method \cite{Schiller00} allows a simultaneous determination of the level density and the $\gamma$-ray 
strength function ($\gamma$SF) from particle-$\gamma$ coincidences. In the nuclear quasi-continuum region, 
these quantities provide information on the average properties of excited states and their decay and branching ratios.

In this work, we report on the level densities below the neutron separation energy for the $^{89,90}$Y isotopes 
with neutron number $N=50$ and 51, respectively. The experimental results are compared with a simple combinatorial model 
from which additional information on parity and spin distribution can be obtained.

In Sect.~II the experimental results are described. The nuclear level densities are extracted in Sect.~III, 
and in Sect.~IV, model predictions of the impact of the $N=50$ closure is discussed. Summary and conclusions are given in Sect.~V.

\section{Experimental results}
\label{sec:exp}

The experiments were performed at the Oslo Cyclotron Laboratory (OCL) 
with a 17-MeV proton beam and an 11-MeV deuteron beam. 
The target was a 2.25 mg/cm$^2$ thick metallic foil of naturally monoisotopic $^{89}$Y.

The charged outgoing particles were measured with the SiRi system~\cite{siri}
comprising 64 $\Delta E - E$ silicon telescopes with thicknesses 
of 130 and 1550 $\mu$m, respectively. The Si detectors were placed in backward direction
covering eight angles from $\theta = 126^\circ$ to $140^\circ$
relative to the beam axis. By setting 2-dimensional gates on the ($\Delta E, E)$ matrix,
outgoing protons could be selected to define the desired $^{89}$Y($p,p' \gamma$)$^{89}$Y
and $^{89}$Y($d,p \gamma$)$^{90}$Y reactions.
The coincident $\gamma $ rays were measured with the CACTUS
array \cite{CACTUS} consisting of 26 collimated $5" \times 5"$ NaI(Tl)
detectors with a total efficiency of $14.1$\% at $E_\gamma = 1.33$ MeV.
    
\begin{figure*}[t]
 \begin{center}
 \includegraphics[clip,width=2\columnwidth]{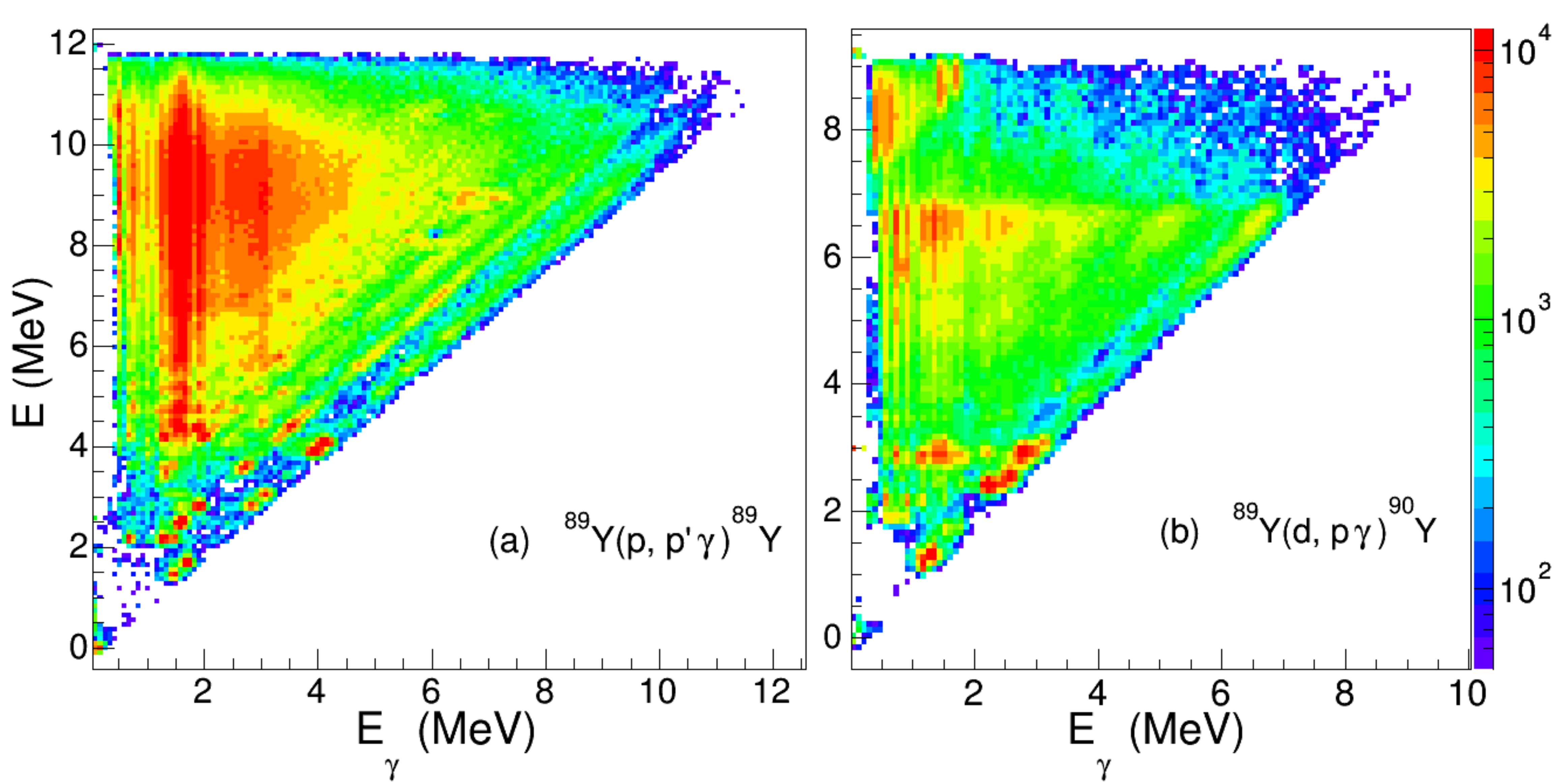}
 \caption{(Color online) Particle-$\gamma$ coincidence matrices for the $^{89}$Y($p, p' \gamma$)$^{89}$Y
 and $^{89}$Y($d, p \gamma$)$^{90}$Y reactions. On the y-axis the measured energy of
 the outgoing particle is used to calculate the initial
 excitation energy $E$ of the residual nucleus. Each NaI spectrum at a given $E$ is unfolded with the NaI
 response function.}
 \label{fig:matrices}
 \end{center}
 \end{figure*}

The energy from the outgoing charged particles can be converted into excitation energy of 
the residual nucleus when states below the neutron separation energy are populated.
Figure~\ref{fig:matrices} shows the particle-$\gamma$ matrices $(E, E_{\gamma})$
for the two reactions with prompt coincidence requirement, where
the $\gamma$ spectra have been unfolded with the NaI response
functions~\cite{gutt1996}. The neutron separation energy
of $^{90}$Y is clearly seen at $E\approx S_n=6.857$ MeV, where the $\gamma$
intensity/multiplicity suddenly drops.

The vertical lines of the coincidence matrices display yrast transitions from the last steps in the $\gamma$-cascades.
Furthermore, we see diagonals, where the $E=E_\gamma$ line represents
primary $\gamma$-rays
that feed directly the ground states of spin/parity 1/2$^{-}$ and 2$^{-}$ in $^{89,90}$Y, respectively.
In $^{89}$Y the second diagonal represents decay to the 9/2$^{-}$ state at 909 keV. The third diagonal is the result of
direct decay to the 3/2$^{-}$(1507 keV) and 5/2$^{-}$(1745 keV) states.
 
The energy distribution of first-generation or primary $\gamma$ rays can be extracted from the unfolded total $\gamma$ spectra
of Figs.~\ref{fig:matrices} (a) and (b). Let $U^E(E_{\gamma})$ be the
unfolded $\gamma$ spectrum at a certain initial excitation energy $E$.
Then the first-generation or primary spectrum can be obtained by a subtraction of a weighted sum of
$U^{E'}(E_{\gamma})$ spectra for $E'$ below  $E$:
\begin{equation}
F^E(E_{\gamma})=U^E(E_{\gamma}) - \sum_{E' < E}w_{E'}U^{E'}(E_{\gamma}).
\end{equation}
The weighting coefficients $w_{E'}$ are determined in an iteration process described in Ref.~\cite{Gut87}.
After a few iterations, the weighting coefficients $w_{E'}$ (as function of $E'$) are equal to the distribution
$F^E(E_{\gamma})$, which is exactly what is expected, namely the primary $\gamma$ spectrum
equals the weighting function.
The subtraction technique is based on the assumption that the decay $\gamma$-energy
distribution is the same whether the levels were populated directly by
the nuclear reaction or by $\gamma$ decay from higher-lying states. 
In particular, this assumption is fulfilled when states have the same 
relative probability to be populated by the two processes, 
since $\gamma$-branching ratios are properties of the levels themselves.

 \begin{figure*}[bt]
 \begin{center}
 \includegraphics[clip,width=1.8\columnwidth]{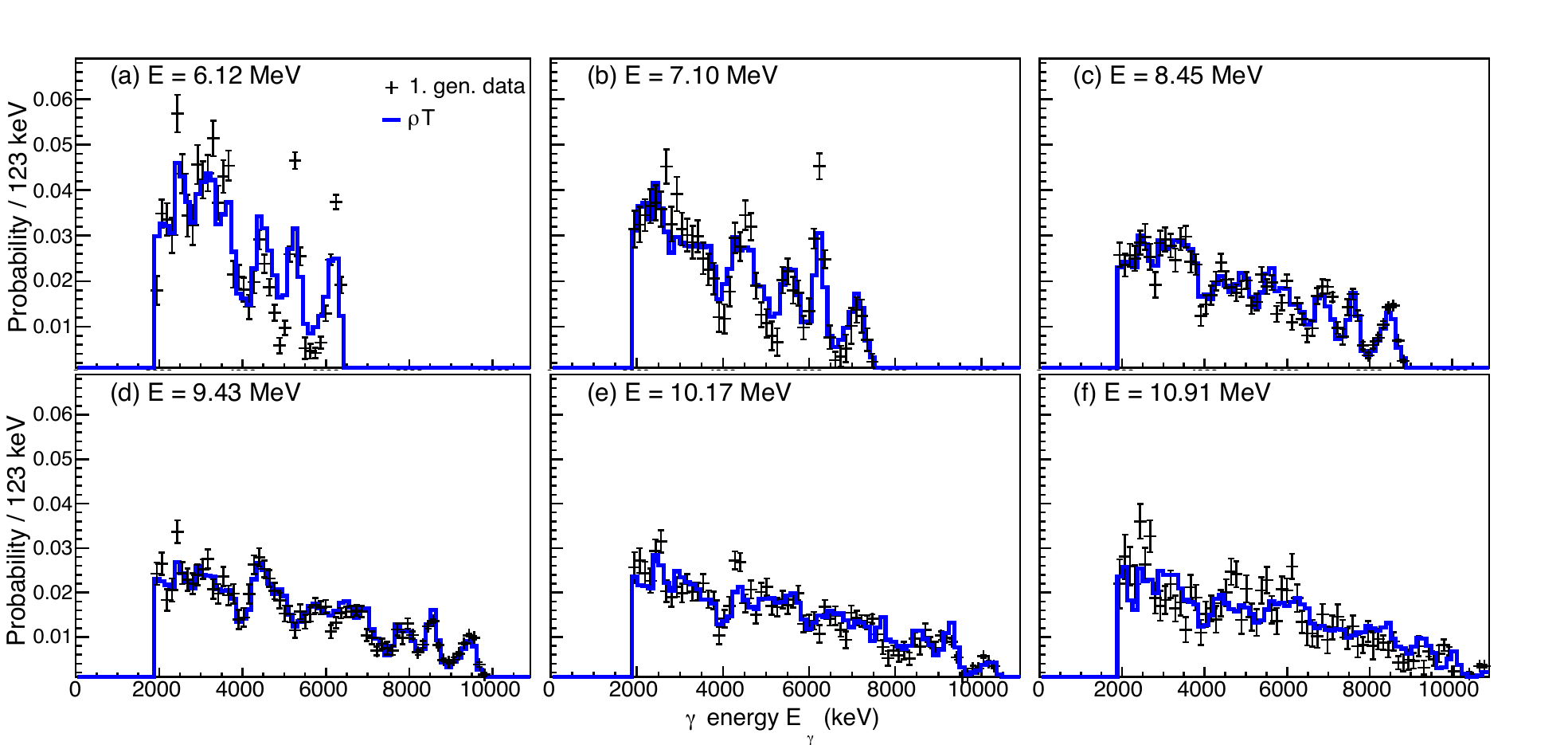}
 \caption{(Color online) First-generation spectra from various initial excitation energies $E$ (crosses). The spectra
 are compared to the product of the level density and transmission coefficient vectors
 i.e.~$\rho(E-E_{\gamma}) {\mathcal{T}}(E_{\gamma})$ (blue lines). 
 All spectra are normalized to unity. Both the $\gamma$
 and excitation energy dispersions are 123 keV/ch.}
 \label{fig:does}
 \end{center}
 \end{figure*}
A matrix $P(E,E_{\gamma})$ covering all initial excitation energies $E$, is obtained
by extracting the primary $\gamma$ spectra $F^E(E_{\gamma})$ for all $E$.
The statistical part of this
landscape of probabilities\footnote{Each $\gamma$ spectrum is normalized
by $\sum_{E_{\gamma}}P(E,E_{\gamma})=1$.} is then assumed 
to be described by the product of two vectors
\begin{equation}
P(E, E_{\gamma}) \propto   \rho(E-E_{\gamma}){\cal{T}}(E_{\gamma}) ,\
\label{eqn:rhoT}
\end{equation}
where the decay probability is proportional to the
level density at the final energy $\rho(E-E_{\gamma})$ according
to Fermi's golden rule~\cite{dirac,fermi}. The decay is also proportional
to the $\gamma$-ray transmission coefficient ${\cal{T}}$, which,
according to the Brink hypothesis~\cite{brink}, is independent of excitation energy;
only the transitional energy $E_{\gamma}$ plays a role. 
    
The relation (\ref{eqn:rhoT}) is a rather strong ansatz that makes it
possible to simultaneously extract the two one-dimensional vectors $\rho$
and ${\cal{T}}$ from the two-dimensional landscape $P$. The justification
of this has been experimentally tested for many nuclei by the Oslo group
and a survey of possible errors in the Oslo method has been discussed
in Ref.~\cite{Lars11}.
    
Before we proceed further,
we must select a part of the $P$ matrix where the primary $\gamma$ spectra are dominantly statistical. For
$^{89}$Y the excitation energy region chosen is 5.4~MeV $< E < $ 11.0~MeV with $E_{\gamma}> 2.0$~MeV, and for
$^{90}$Y we choose                           4.0~MeV $< E < $  6.8~MeV with $E_{\gamma}> 1.5$~MeV.
With these cuts in the matrices, we use the iteration procedure of Schiller {\em et al.}~\cite{Schiller00}
to determine $\rho$ and ${\cal{T}}$ by a least $\chi ^2$ fit using relation (\ref{eqn:rhoT}).

The applicability of relation (\ref{eqn:rhoT}) and the quality of the fitting
procedure are demonstrated in Fig.~\ref{fig:does}. The agreement is very satisfactory when one
keeps in mind that the $\gamma$-decay pattern fluctuates from level to level.
With the rather narrow excitation energy bins of 123 keV, each $\gamma$ spectrum will be subject to significant
Porter-Thomas fluctuations~\cite{PT} responsible for local deviations for individual primary spectra
compared to the global average $\rho{\cal{T}}$.
\section{Normalization of the level density}

The functional form of $\rho$ and $\cal{T}$ are uniquely identified through the fit, 
but the scale and slope of these functions are still undetermined.
It is shown in Ref.~\cite{Schiller00} that functions generated by the transformations:
\begin{eqnarray}
\tilde{\rho}(E-E_\gamma)&=&A\exp[\alpha(E-E_\gamma)]\,\rho(E-E_\gamma),
\label{eq:array1}\\
\tilde{{\mathcal{T}}}(E_\gamma)&=&B\exp(\alpha E_\gamma){\mathcal{T}} (E_\gamma)
\label{eq:array2}
\end{eqnarray}
give identical fits to the primary $\gamma$ spectra, as the examples shown in Fig.~\ref{fig:does}.
In the following, we will estimate the parameters $A$ and $\alpha$ from systematics and other experimental data. 
The normalization of ${\mathcal{T}}$ by the
constant $B$, only concerns the $\gamma$SF that will not be discussed in the present work.
    
The standard approach to find $A$ and $\alpha$ is to reproduce the level density where
one assumes that a complete level scheme is known, and to fit to the level density
extracted from average neutron resonance capture spacing $D_0$ at the neutron separation energy $S_n$.

Unfortunately, there are no experimental $D_0$ values for the $N=50$ isotope $^{89}$Y, 
since  $^{88}$Y is unstable. We will therefore investigate the known $D_0$s in this mass region  and corresponding
level densities $\rho(S_n)$,  and compare with the systematics evaluated in order
to estimate $\rho$ for $^{89}$Y at high excitation energy.

 \begin{figure}[ht]
 \begin{center}
 \includegraphics[clip,width=\columnwidth]{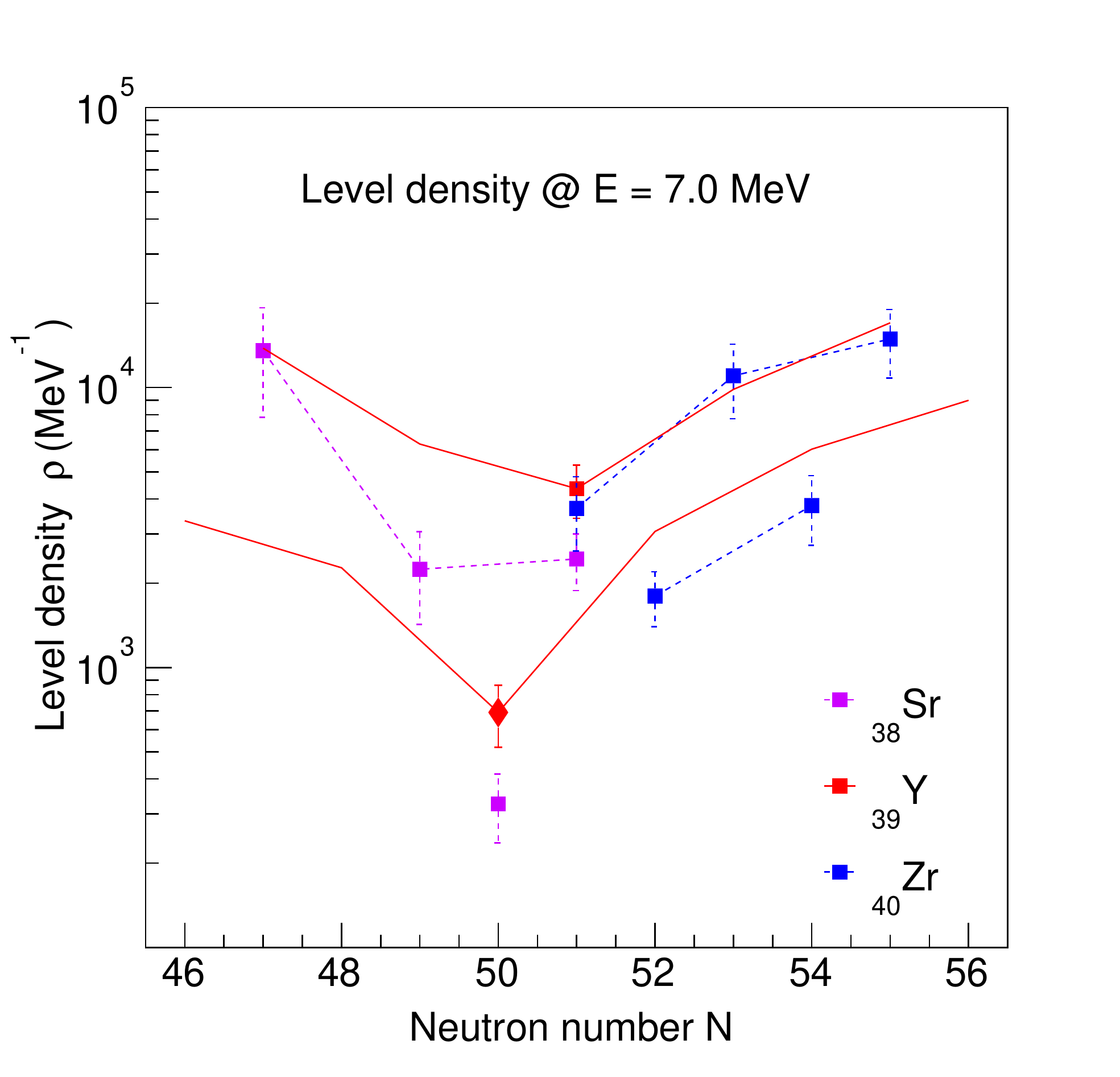}
 \caption{(Color online) Level densities at $E=7$ MeV (square data points) 
 extracted from neutron resonance spacings $D_0$~\cite{RIPL3}. 
 The data points of even and odd neutron numbers $N$ are connected
 by dashed lines. The red solid lines are from the global systematics of Ref.~\cite{egidy2009}.
 The red diamond point is used for normalizing the level density of $^{89}$Y.}
 \label{fig:y_syst}
 \end{center}
 \end{figure}

The level density at the neutron separation energy $\rho(S_n)$ is calculated from the
$\ell = 0$ neutron resonance spacings $D_0$  assuming a spin distribution~\cite{Ericson}
\begin{equation}
g(E,I) \simeq \frac{2I+1}{2\sigma^2}\exp\left[-(I+1/2)^2/2\sigma^2\right],
\label{eq:spindist}
\end{equation}
using $E=S_n$  and $I$ is the spin. The spin cut-off parameter $\sigma$ at $S_n$
is taken from Ref.~\cite{egidy2009}.
    
The $N=50$ isotones have almost twice the $S_n$ value as the other neighboring nuclei.
Therefore, in order to place the data points on the same footing, we use $\rho(S_n)$ from known level spacings $D_0$ and the
constant-temperature level density formula~\cite{Ericson}
\begin{equation}
\rho(E)=\frac{1}{T}\exp\left({\frac{E-E_0}{T}}\right)
\label{eq:ct}
\end{equation}
to estimate new anchor points at a common excitation energy of $E=7$~MeV for all the considered nuclei. Here, we use
the parameter $T$ from Table II of Ref.~\cite{egidy2009} and energy shift $E_0$ so as to reproduce $\rho(S_n)$. 
Figure~\ref{fig:y_syst} shows the deduced level densities
for the Sr, Y and Zr isotopes for which experimental $D_0$ values are available~\cite{RIPL3}.
    
The $\rho({\rm 7 MeV})$ calculations of Fig.~\ref{fig:y_syst} clearly reveal a lower level density as one is approaching
the $N=50$ shell gap. Also the even-$N$ isotopes have several times lower level density as their odd-$N$ neighbors.
The most important anchor points to estimate the level density for $^{89}$Y, is the Sr points at $N=$ 49, 50 and 51 together
with the $^{90}$Y point at $N=51$. The red lines are estimates of yttrium isotopes based
on systematics with global parametrization~\cite{egidy2009}, but scaled with a factor 0.18 in order to match the
$\rho({\rm 7 MeV})$ point of  $^{90}$Y. In this way, we find an estimate for the level density of $^{89}$Y of 
$\rho(7 {\rm MeV})= 690 \pm 170$~MeV$^{-1}$ (marked with a diamond). This value is also supported by the 
fact that the ratios $\rho(^{89}$Sr$_{51})/\rho(^{88}$Sr$_{50})\approx \rho(^{90}$Y$_{51})/\rho(^{89}$Y$_{50})$, 
as found in Fig.~\ref{fig:y_syst}, which is expected if the $N=50$ gap in Sr and Y is roughly the same.

\begin{figure}[bt]
 \begin{center}
 \includegraphics[clip,width=\columnwidth]{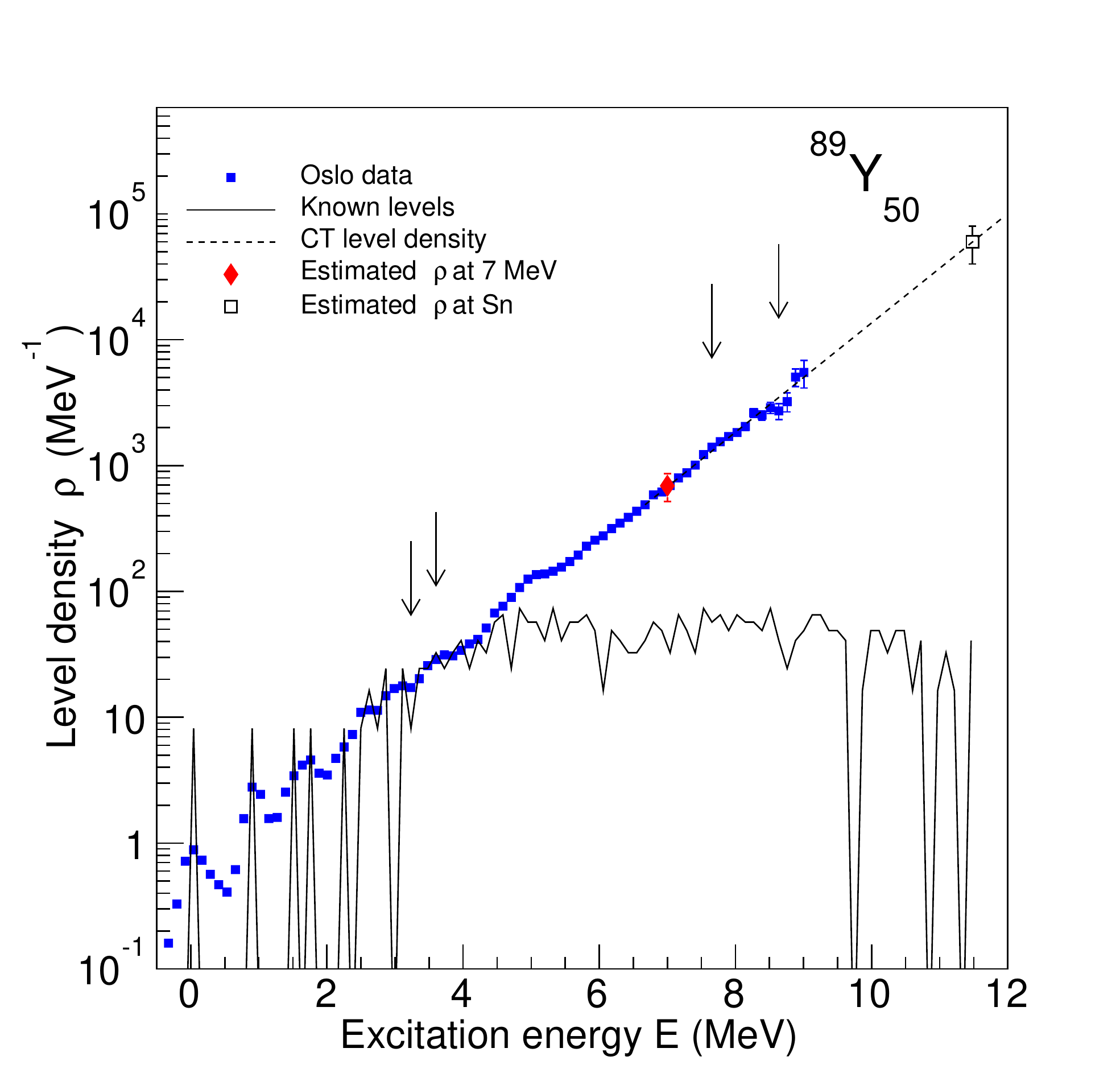}
 \caption{(Color online) Normalization of the nuclear level density (filled squares) of $^{89}$Y. 
 At low excitation energies, the data are normalized (between the
arrows) to known discrete levels (solid line). At higher excitation energies, 
the data are normalized to the constant-temperature level density (dashed line) going through
the point $\rho(7 {\rm MeV})$ (red diamond), which is estimated from the systematics of Fig.~\ref{fig:y_syst}.
The $\rho$ at $S_n$ is determined by extrapolation with a constant-temperature level density with
parameters from Table~\ref{tab:parameters}.
}
 \label{fig:counting_y89}
 \end{center}
 \end{figure}

\begin{figure}[bt]
 \begin{center}
 \includegraphics[clip,width=\columnwidth]{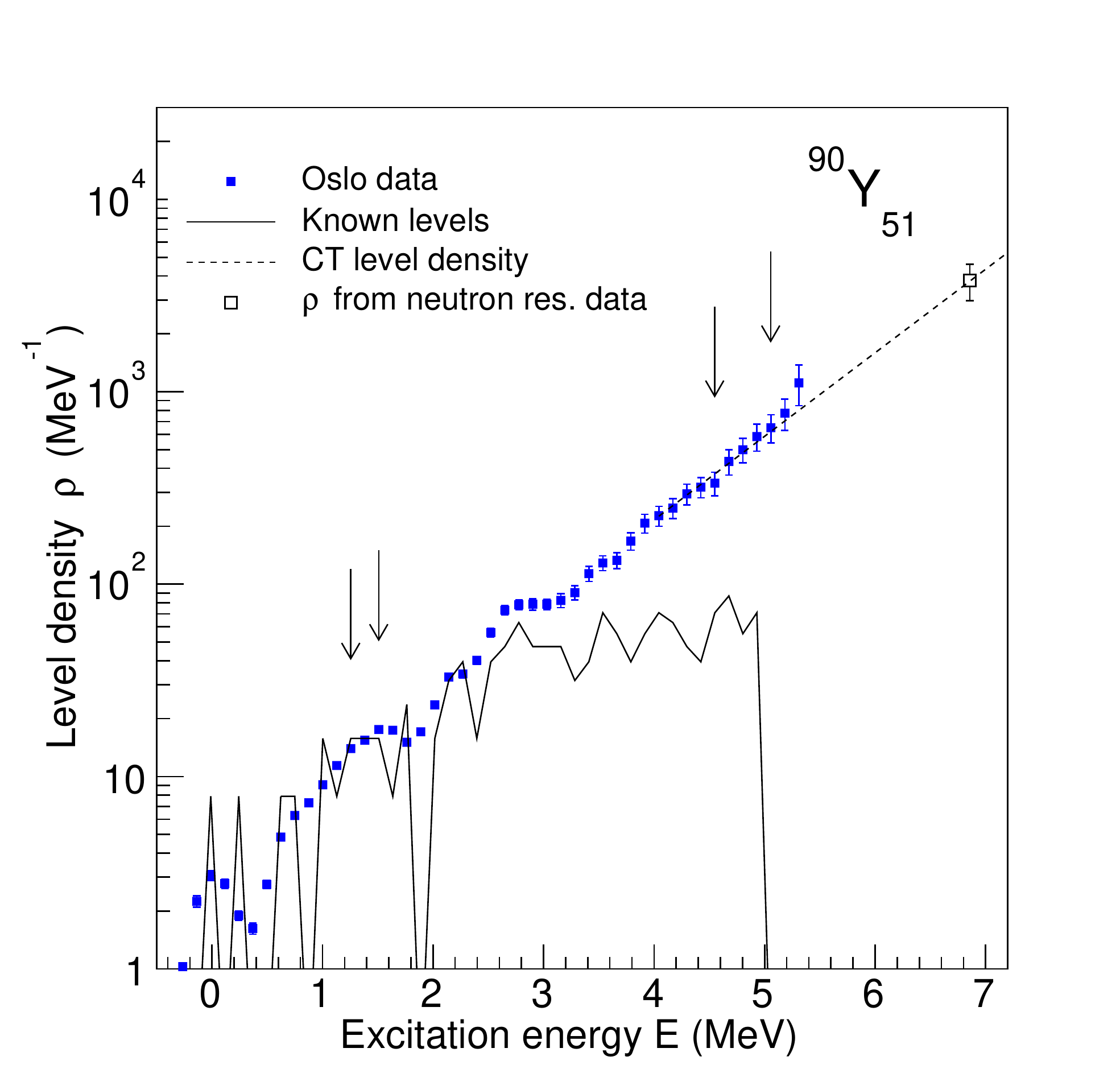}
 \caption{(Color online) The level density of $^{90}$Y (see text of Fig.~\ref{fig:counting_y89}).
The data point at $S_n$  is calculated from the neutron resonance spacing $D_0$ of Table~\ref{tab:parameters}.}
 \label{fig:counting_y90}
 \end{center}
 \end{figure}

The low-energy level schemes of $^{89,90}$Y are fairly well known up to a level density of $\rho \approx 30$ MeV$^{-1}$ as
shown by solid lines in Figs.~\ref{fig:counting_y89} and \ref{fig:counting_y90}. Thus, this information gives a reliable 
normalization at low excitation energy. At higher energy, we use the value extracted from Fig.~\ref{fig:y_syst} at 
$E=7$ MeV for $^{89}$Y and the level density extracted from the $D_0$ value for $^{90}$Y.
A summary of the data used for the normalizations are listed in Table~\ref{tab:parameters}.

 \begin{table}[htb]
    \caption{Parameters used for the normalizations of the level density.}
    \begin{tabular}{c|c|ccccc}
    \hline
    \hline
    Nucleus   &$S_n$&   $T$  & $E_0$   &$\sigma(S_n)$&  $D_0$   &   $\rho(S_n)$ \\
              &(MeV)&(MeV)   &(MeV)    &             &  (eV)    & ($10^3$MeV$^{-1}$)\\
    \hline
    $^{89}$Y&11.478&   1.00  &  0.355  &    3.60  & 106(35)$^a$    & 60(20)$^a$  \\
    $^{90}$Y& 6.857&   1.00  & -0.648  &    3.61  & 3700(400)$^b$ & 3.77(81) \\
    

    \hline
    \hline
    \end{tabular}
    \\$^a$) Estimated from the diamond data point at $E=7$ MeV in Fig.~\ref{fig:y_syst}.
    \\$^b$) From Ref.~\cite{RIPL3}.
    \label{tab:parameters}
    \end{table}

\section{Model description and comparison with data}
The experimental level densities are shown in Figs.~\ref{fig:counting_y89} and \ref{fig:counting_y90}. At
low excitation energy our data agree very well with the detailed structures found from known discrete levels. Above $\approx$ 3 MeV
the level densities follow closely the constant-temperature formula of Eq.~(\ref{eq:ct}). Since the temperature in the 
microcanonical ensemble
is given by $T(E)=d\ln\rho(E)/dE \approx$ const., we actually observe a system which keeps the same temperature even 
when the intrinsic energy increases. This is the ultimate  sign of a first-order phase transition,  
as further elaborated in Ref.~\cite{luciano2014}.

In order to microscopically describe the level density at high excitation energy, detailed knowledge of the
nucleon-nucleon matrix elements is not necessary since only average properties are of
interest. Previously it has been shown~\cite{gutt2001} that the essential mechanism for increasing level density
is to break $J=0$ nucleon pairs, giving 25 - 35 more levels for each pair broken. Thus, a simple model
has been developed, which includes these most important features~\cite{scand2007}. It is also appreciable that the model
uses the microcanonical ensemble with a fixed energy, pressure and volume, and does not rely on an infinitely large heat bath
 (canonical ensemble). The termal contact with such a reservoir is conceptually difficult to apply for an isolated system
 like the nucleus. In the following, we will
 use the abbreviation $\mu$CM for our microcanonical combinatorial model.
 
 \begin{table}[htb]
    \caption{Most prominent proton ($\lambda_{\pi}= 44.708$~MeV) and
    neutron ($\lambda_{\nu}=50.510$~MeV) Nilsson orbitals for $^{89}$Y.}
    \begin{tabular}{c|c|ccccc}
    \hline
    \hline
    Orbital   &$e_{\rm sp}$& \multicolumn{5}{c}{Spherical $j$ component} \\
    $\Omega^{\pi}[Nn_z\Lambda]$ &(MeV)& $1/2$ & $3/2$ & $5/2$  & $7/2$  & $9/2$\\
    \hline
   {\bf Protons}&&&&&&\\
   1/2$^-$[310]& 42.394&  -0.2335&-0.4345& 0.8672& -0.0687& \\
   3/2$^-$[312]& 42.489&   0.0000&-0.8225&-0.5449& -0.1634& \\
   3/2$^-$[301]& 43.256&   0.0000&-0.5482& 0.8359& -0.0280& \\
   5/2$^-$[303]& 43.972&   0.0000& 0.0000& 0.9979&  0.0652& \\
   1/2$^-$[301]& 44.412&  -0.9013&-0.2337&-0.3628& -0.0371& \\
   1/2$^+$[440]& 46.876&   0.0374& 0.0144&-0.2327& -0.0170& 0.9716 \\
   3/2$^+$[431]& 47.147&   0.0000& 0.0162&-0.1850& -0.0413& 0.9818 \\
   5/2$^+$[422]& 47.641&   0.0000& 0.0000&-0.1119& -0.0517& 0.9924 \\
   7/2$^+$[413]& 48.318&   0.0000& 0.0000& 0.0000& -0.0467& 0.9989 \\
   9/2$^+$[404]& 49.147&   0.0000& 0.0000& 0.0000&  0.0000& 1.0000 \\
   1/2$^+$[431]& 52.323&   0.3428& 0.1933&-0.8794& -0.1391&-0.2291 \\
   \hline
   {\bf Neutrons}&&&&&&\\
   1/2$^+$[440]& 46.876&  0.0374&  0.0144& -0.2327& -0.0170&  0.9716\\
   3/2$^+$[431]& 47.147&  0.0000&  0.0162& -0.1850& -0.0413&  0.9818\\
   5/2$^+$[422]& 47.641&  0.0000&  0.0000& -0.1119& -0.0517&  0.9924\\
   7/2$^+$[413]& 48.318&  0.0000&  0.0000&  0.0000& -0.0467&  0.9989\\
   9/2$^+$[404]& 49.147&  0.0000&  0.0000&  0.0000&  0.0000&  1.0000\\
   1/2$^+$[431]& 52.323&  0.3428&  0.1933& -0.8794& -0.1391& -0.2291\\
   3/2$^+$[422]& 53.145&  0.0000&  0.1831& -0.9315& -0.2509& -0.1891\\
   1/2$^+$[420]& 53.659& -0.0493& -0.4027& -0.2384&  0.8817& -0.0338\\
   5/2$^+$[413]& 54.275&  0.0000&  0.0000& -0.9358& -0.3305& -0.1227\\
   3/2$^+$[411]& 54.323&  0.0000& -0.1823& -0.2868&  0.9404& -0.0115\\
   5/2$^+$[402]& 55.143&  0.0000&  0.0000& -0.3343&  0.9424&  0.0114\\
    \hline
    \hline
    \end{tabular}
    \label{tab:nilsson}
    \end{table}

The Nilsson model~\cite{nilsson1955} is applied to generate a set of single-particle
orbitals. The Nilsson parameters used are: the quadrupole deformation
$\epsilon_2=0.1$, the spin-orbit splitting $\kappa = 0.066$ and the centrifugal parameter $\mu =  0.32$.
The harmonic-oscillator quantum number is taken as $\hbar\omega_{\rm osc}=41A^{-1/3}$. In order to obtain a reasonable
description, the gaps obtained from the ($\epsilon_2, \kappa, \mu$) parameters had to be
increased by 1.0 and 1.5 MeV at the $N/Z= 40$  and 50 gaps, respectively.
The same parameter set is used for protons and neutrons. The most prominent orbitals with their $j$-component
are shown in Table~\ref{tab:nilsson}.
 \begin{figure}[bt]
 \begin{center}
 \includegraphics[clip,width=\columnwidth]{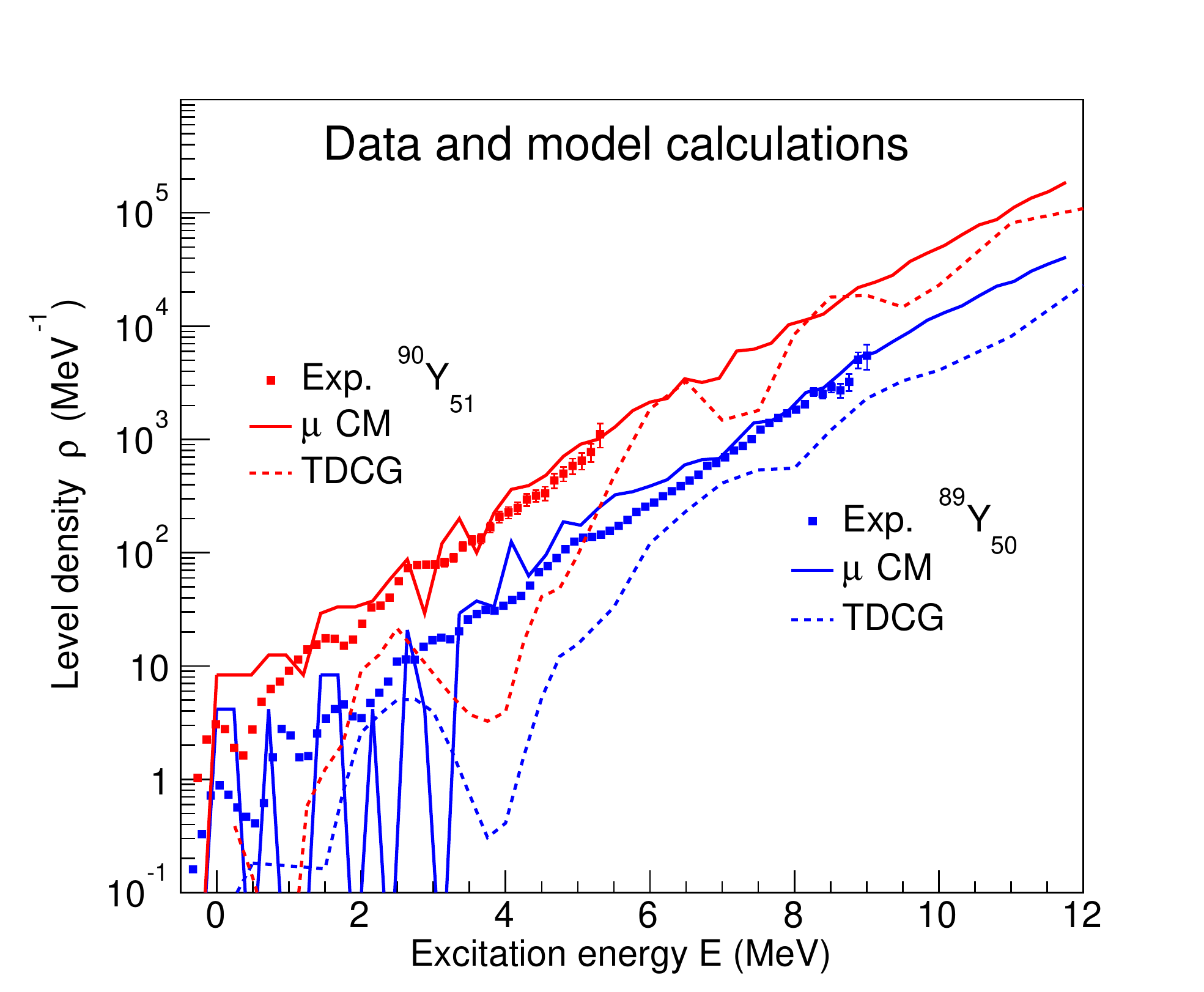}
 \caption{(Color online) Experimental level densities for $^{89,90}$Y (data points) compared with models. 
 The solid lines are predictions of the $\mu$CM. The dashed lines are TDCG calculations of Hilaire {\em et al.}~\cite{hilaire2012}.}
 \label{fig:combi}
 \end{center}
 \end{figure}
 \begin{figure}[bt]
 \begin{center}
 \includegraphics[clip,width=\columnwidth]{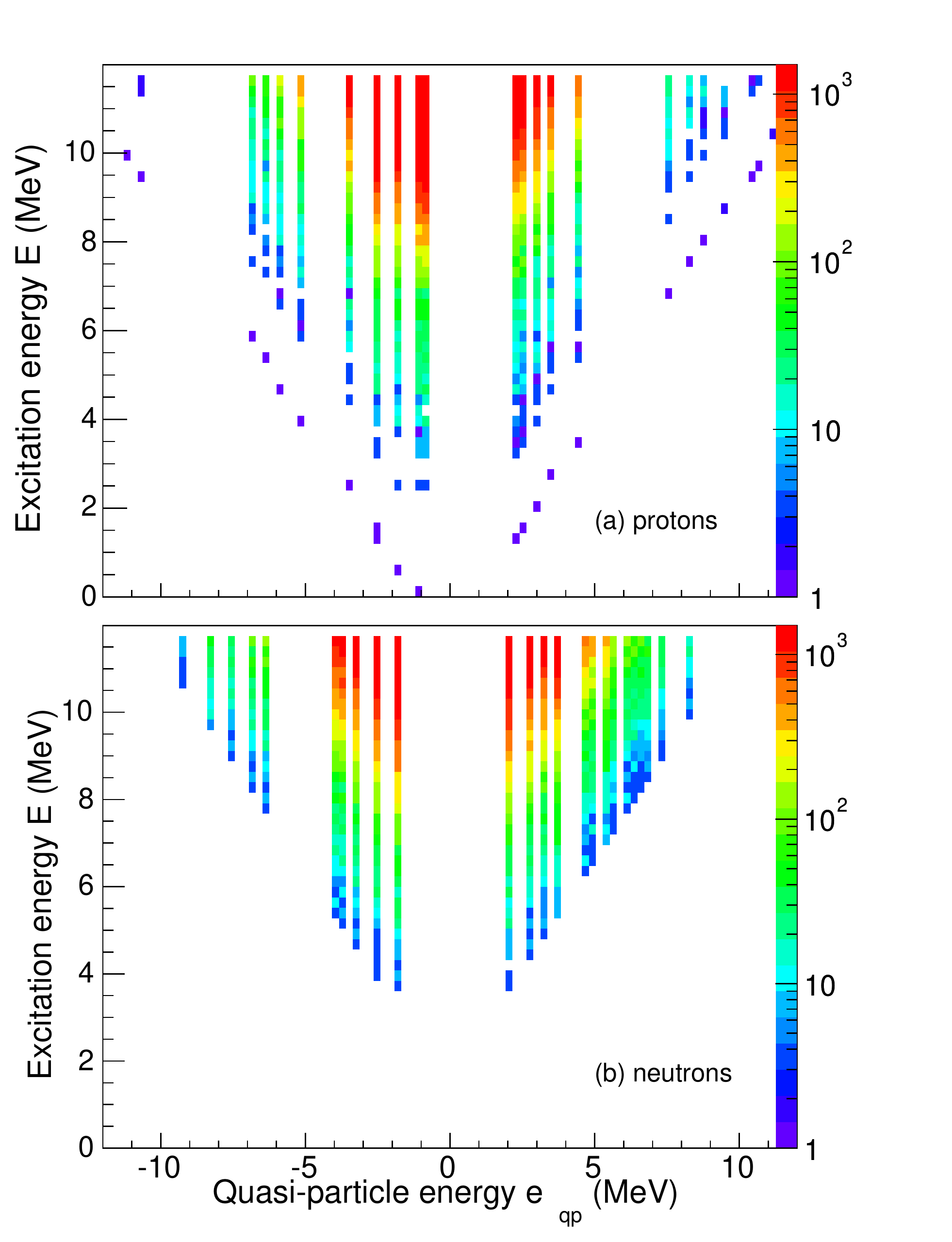}
 \caption{(Color online) Active proton (a) and neutron (b) quasi-particles at various excitation energies $E$ in $^{89}$Y.
 The quasi-particle energies $e_{\rm qp}$ have been assigned positive and negative values above and below the Fermi surface, respectively. 
 The z-axis (shown in colors) tells what orbitals are most active at a certain excitation energy.}
 \label{fig:orbitnp}
 \end{center}
 \end{figure}
    
The Nilsson orbital energies $e_{\rm sp}$ for an axially deformed nuclear shape can now be transferred to
single quasi-particle energies by the Bardeen-Cooper-Schrieffer (BCS) theory~\cite{BCS}:
\begin{equation}
e_{\rm qp} = \sqrt{(e_{\rm sp}-\lambda)^2 + \Delta^2}, \
\label{eqn:qp}
\end{equation}
where the Fermi level $\lambda$ is determined by the number of protons or neutrons. For the
pairing energy parameter we use $\Delta_{\pi} =1.5$ MeV and $\Delta_{\nu} =2.0$ MeV in reasonable agreement with the 
odd-even mass differences in this mass region.
        
In order to obtain the number of levels per MeV, we combine all possible proton and neutron
quasi-particles giving an energy sum less than an upper excitation energy $E$, in our case 
up to $E=12$~MeV. This means that
we include orbitals from 12 MeV below the Fermi level up to 12 MeV above the Fermi level. 
For example  we include 31  quasi-proton and 37 quasi-neutron orbitals for the $^{90}$Y nucleus.
        
The number of levels $N(E)$ at the excitation energy $E$ is incremented each time a combination of quasiparticles 
within a bin size of $\Delta E = 0.24$~MeV fulfills:
\begin{equation}
E = \sum_{{\Omega_\pi^\prime,\Omega_\nu^\prime}}  e_{\rm qp}(\Omega_\pi^\prime) + e_{\rm qp}(\Omega_\nu^\prime) + V ,
\label{eqn:qpe}
\end{equation}
where $\Omega_\pi$ and $\Omega_\nu$ are the angular momentum projections of protons and neutrons onto the symmetry axis. 
When all possible combinations of quasi-particles have been performed, the level density is finally given by $\rho(E)= N(E) / \Delta E$.
        
Each Nilsson orbital is doubly degenerated, i.e.~$\Omega$ and $-\Omega$ orbitals have the same energy. In cases
when two or more $\Omega$s are combined, one would get degenerate states by time reversal of one or more $\Omega$s. As an example, a
three quasi-particle state can be found in these configurations
    $(\Omega_1,\Omega_2, \Omega_3)$,
    $(-\Omega_1,\Omega_2, \Omega_3)$,
    $(-\Omega_1,-\Omega_2, \Omega_3)$,
    $(\Omega_1,-\Omega_2, \Omega_3)$ and
    $(\Omega_1,-\Omega_2, -\Omega_3)$
giving all together five configurations with a different sums of angular momenta $j$. 
In order to prevent such a bunching of states at the same excitation energy, 
we have added a residual interaction $V$ by a random Gaussian distribution with an average absolute energy of 100 keV.

The most important Nilsson orbitals for $^{89}$Y (and very similar for $^{90}$Y) are listed in Table~\ref{tab:nilsson}. 
The main components of these orbitals are proton $p_{1/2}$, $p_{3/2}$, $f_{5/2}$, $f_{7/2}$ and $g_{9/2}$ and neutron 
$g_{9/2}$, $d_{5/2}$ and $d_{7/2}$ spherical states. The proton 1/2$^-$[301] orbital at $e_{\rm sp}=44.412$~MeV 
is closest to the Fermi level and becomes the ground state with $I^{\pi}=1/2^-$. It has a large $j=1/2$ component 
originating from the spherical $\pi p_{1/2}$ state.
    
Figure~\ref{fig:combi} shows that the $\mu$CM level densities  describe satisfactorily the experimental data points for 
both $^{89,90}$Y. This gives confidence to our simple model and that it is possible to draw some general conclusions 
on certain main structural properties in the quasi-continuum. 

The densities at low excitation energy are well reproduced and indicate that the densities of Nilsson orbitals are 
realistic and thus verify the size of the shell gaps. One should note that there has been made no effort to reproduce 
the detailed ordering of the low-lying states. At high excitation energy, the calculations also reproduce rather well 
the general trend of a constant-temperature level density with a critical temperature of $T\approx 1.0$~MeV. 
The level density of the odd-odd $^{90}$Y is on the average 6 times the level density of the even-odd $^{89}$Y nucleus. 
This also holds for the lowest excitation region where there are 36 known levels below $E=2.6$~MeV in $^{90}$Y and 
only 7 levels in $^{89}$Y, see Ref.~\cite{NNDC}. Due to the large $N=50$ gap, the major part of the levels are proton 
states generated by orbitals around the $Z=39$ Fermi level. Thus, for $^{90}$Y the last valence neutron outside the 
$N=50$ gap generates many new states, indicating that there are several single-neutron orbitals available above this gap. 
This feature is also verified in the Nilsson calculations, see Table~\ref{tab:nilsson}.
    
We also compare our data in Fig.~\ref{fig:combi} with the temperature-dependent combinatorial level densities with 
the D1M Gogny force of Hilaire {\em et al.}~\cite{hilaire2012} (dashed lines marked with TDCG). The TDCG level densities 
are rather well behaving above $E\approx 6$~MeV. However, at lower excitation energies the TDCG predictions are a 
factor $10-100$ lower than the experimental data. This becomes particularly clear for both $^{89,90}$Y at $É\approx4$~MeV, 
where the known level density from counting are almost 100 times higher than predicted with the 
TDCG calculations~\footnote{An adjustment of the proposed $\delta$-shift for the TDCG level densities~\cite{hilaire2012} 
would moderately improve the agreement.}. The TDCG results demonstrates how difficult it is to get good agreement 
in the vicinity of closed shells, and that these calculations probably are dealing with a too large effective $N=50$ shell gap.
        
One may ask if it is necessary to include all proton and neutron orbitals up to $|e_{\rm qp}|< E$ in order to describe 
the level density at $E$, as performed for the $\mu$CM calculations. This is an adequate question for large shell-model 
calculations where it is unachievable to include so many orbitals as in the present case\footnote{The $\mu$CM computer 
code consumes maximum 4 min. of CPU time in this mass region.}. Figure~\ref{fig:orbitnp} shows the quasi-particles that 
participate at a certain excitation energy $E$. The colors (z-axis) give how many times a certain quasi-particle with 
energy $e_{\rm qp}$ is included in the wave functions at $E$ within an energy bin of 240 keV. For example at $E=12$~MeV 
the proton orbital $1/2^-[301]$ appears 4505 times, whereas the deeply lying $3/2^+[202]$ only appears 2 times. 
Thus, at high excitation energy the many quasi-particle configurations composed of orbitals close to the Fermi level are 
responsible for the main part of the level density.
        
The protons are seen to be responsible for the low-lying single-particle regime below $E=3-4$~MeV. For the highest energies 
it is obvious that the orbitals closest to the Fermi level most frequently participate in the wave functions. 
The quasi-particles more than 5 MeV from the Fermi level give a significantly less contribution to the level density, 
and might be truncated. A test where only quasi-particles with $|e_{\rm qp}| < 5$~MeV (instead of 12 MeV) are included 
gives in total a reduction from 60 to 28 active orbitals that corresponds to 10 times shorter CPU time needed. 
The level density becomes $\rho(12 {\rm MeV}) =35888$ MeV$^{-1}$ for the truncated basis, compared to 40545 MeV$^{-1}$ for 
the full basis, which is a rather acceptable reduction.
 \begin{figure}[bt]
 \begin{center}
 \includegraphics[clip,width=\columnwidth]{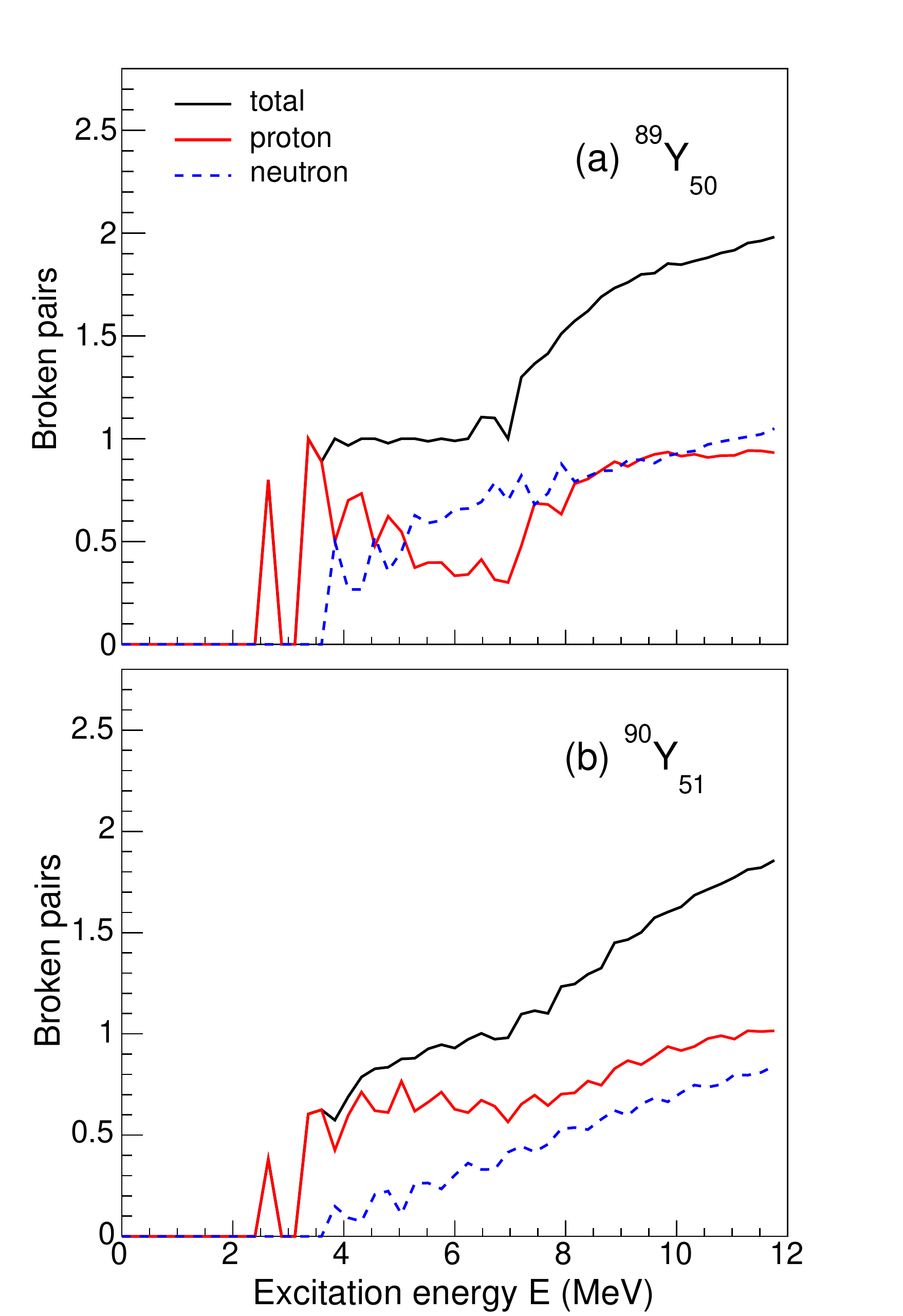}
 \caption{(Color online) Calculated number of proton and neutron pairs broken as function of excitation energy in (a) $^{89}$Y and (b) $^{90}$Y.}
 \label{fig:pairs}
 \end{center}
 \end{figure}
 \begin{figure}[ht]
 \begin{center}
 \includegraphics[clip,width=\columnwidth]{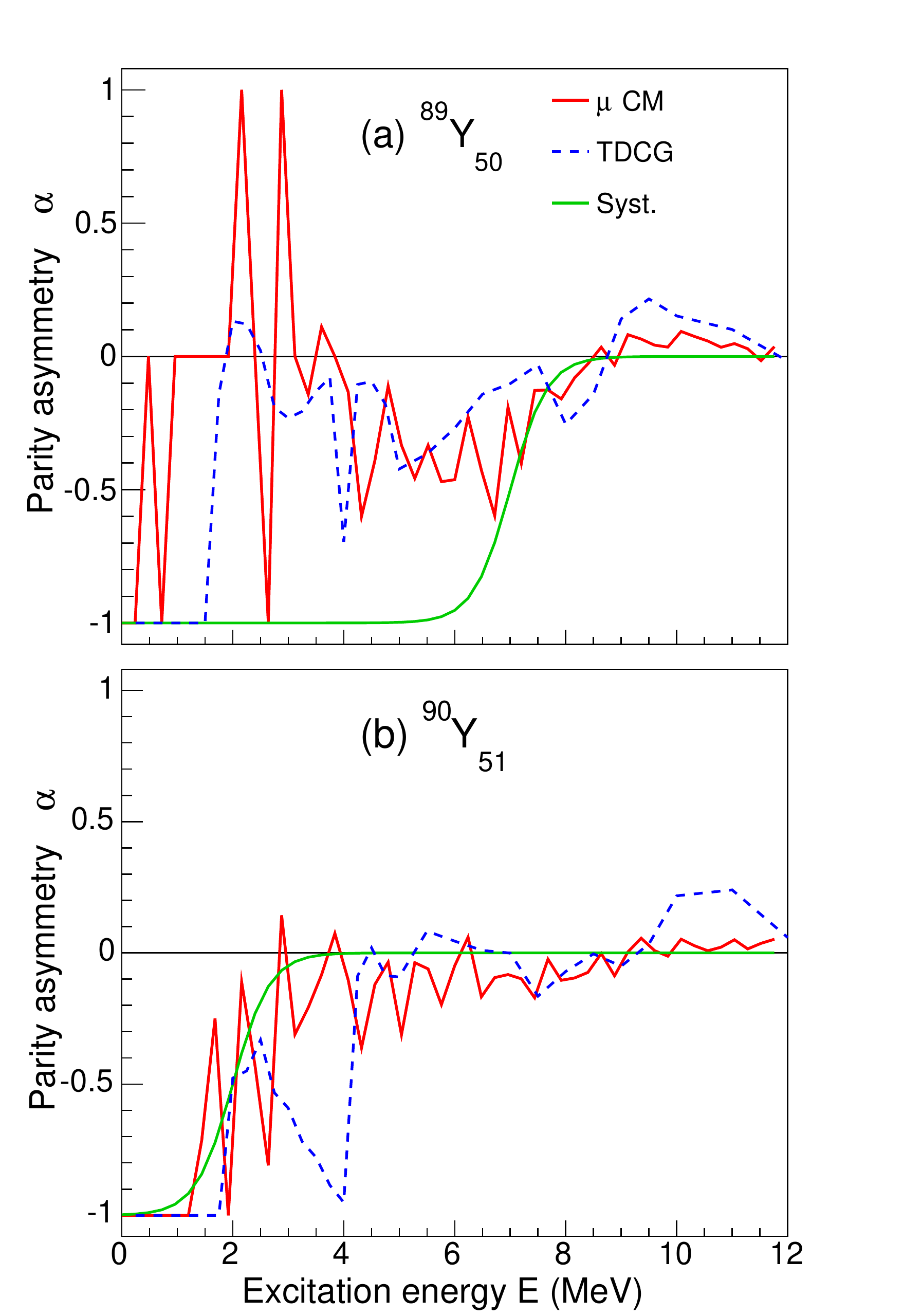}
 \caption{(Color online) Calculated parity asymmetry $\alpha$ as function of excitation energy in (a) $^{89}$Y and (b) $^{90}$Y. 
 The solid red lines are predictions from the $\mu$CM, and the dashed blue lines are TDCG calculations of Hilaire {\em et al.}~\cite{hilaire2012}. 
 The empirical formula based on systematics (green solid line) is taken from Al-Quraishi {\em et al.}~\cite{Ohio2003}, see text.}
 \label{fig:parity}
 \end{center}
 \end{figure}

As already mentioned, the by far most efficient way to increase the level density is by breaking $J=0$ pairs. 
Figure~\ref{fig:pairs} shows that the first pairs are broken at around 3 MeV of excitation energy due to the protons 
in the vicinity of the Fermi level. The $^{89}$Y nucleus experiences an increased contribution of the breaking of neutron pairs 
at 4 MeV and this adds up to one broken pair on the average from $3-7$ MeV of excitation energy. Then the breaking of the next 
proton pair comes into play, giving on average two broken pairs at $E=12$~MeV. For the odd-odd $^{90}$Y isotope the situation 
is different. The breaking of the neutron pair is in some cases blocked by the single-neutron valence particle. 
As a result, the total number of broken pairs is monotonically increasing with excitation energy, and finally reaches the value 
of two broken pairs at 12 MeV. It is interesting to see that both nuclei have one proton and one neutron pair broken at the highest energies. 
Thus, $^{89,90}$Y have in total five and six active quasi-particles at $12$~MeV, respectively.

The spin and parity distributions of nuclear states in the quasi-continuum are determined by the available quasi-particles,
where Table~\ref{tab:nilsson} display the most important ones. The average spin $\left<I\right>$ for $^{89}$Y is rather constant, 
increasing from 3.2 to 3.4 when the excitation energy goes from 7 to 12 MeV. This is consistent with a constant spin-cut off parameter 
of $\sigma = 3.6$ from the systematics of Ref.~\cite{egidy2009}. However, the present combinatorial distribution gives
higher relative intensities for the lower spin values ($I=0, 1, 2$) than predicted by the standard distribution of Eq.~(\ref{eq:spindist}).

The parity distribution is very much controlled by the few negative parity orbitals from the $N_{\rm osc}=3$ oscillator shell, 
see Table~\ref{tab:nilsson}. This hints at an average parity asymmetry in the quasi-continuum region. However, in some cases 
a few parity-intruder states may induce full parity mixing in the many quasi-particle region. This will be investigated in more detail 
in the following where we use the parity-asymmetry parameter to study the parity distribution as function of excitation energy~\cite{asymmetry}:
\begin{equation}
\alpha = \frac{\rho^+-\rho^-}{\rho^++\rho^-},
\end{equation}
where $\rho^+(\rho^-)$ is the density of positive (negative) parity states.

Figure \ref{fig:parity} shows that negative parity states generally dominates over the positive states ($\alpha < 0$),
although there are more positive parity orbitals in the vicinity of the Fermi level. This demonstrates that actual calculations 
have to be performed before any conclusion on the parity distribution can be drawn. In the case of $^{89}$Y the parity asymmetry 
fluctuates dramatically as a result of the few levels at low energy. However, approximately equal number of positive and negative
parity states ($\alpha \approx 0)$ appears above 8 MeV. For $^{89}$Y, the extra valence neutron makes according to the $\mu$CM
calculations $\approx5$ times more states and thus a smoother $\alpha$-curve as function of excitation energy.
The TDCG calculations~\cite{hilaire2012}  (blue dashed lines in Fig.~\ref{fig:parity}) show a parity-asymmetry than
in general follows our $\mu$CM, even as low as $E\approx4$~MeV where we know that the TDCG model severely underestimates the level density.

From a systematical study, Al-Quraishi {\em et al.}~\cite{Ohio2003} have proposed a parity distribution,
which depends on the pairing and/or shell gap parameter $\delta$. From their study, the parity asymmetry is given by
\begin{equation}
\alpha= \pm\frac{1}{1+\exp[(E-\delta)3{\rm MeV^{-1}}]},
\label{eq:grimes}
\end{equation}
where $+$ is used for nuclei where $\alpha$ approaches $+1$ at low $E$ and $-$ if they approach $-1$. Of course, this smooth
function is not appropriate for the low-energy part of $^{89}$Y revealing erratic fluctuations in $\alpha$. However, at high
excitation energies using the values $\delta=7$ and 2 MeV for $^{89,90}$Y, respectively, the empirical formula (green lines in
Fig.~\ref{fig:parity}) describes rather well the $\mu$CM results. 

Within the $\mu$CM model, it seems clear that both nuclei have achieved equally many positive and negative parity states at 
their respectively neutron separation energies. This agrees with actual measurements by Kalmykov {\em et al.}~\cite{parity2007} 
on the neighboring $^{90}$Zr. Between excitation energies of 8 and 11 MeV they find the number of $2^+$ and $2^-$ states to be 
consistent with $\alpha\approx 0$.

\section{Summary and conclusions}
The level densities of $^{89,90}$Y have been extracted and normalized according to the Oslo method. As a consequence of the 
large $N=50$ shell gap, the $^{89}$Y nucleus reveals a very low level density and extremely high neutron separation energy. 
Both nuclei show a constant-temperature level density curve for $E>3$~MeV. The constant-temperature level density behavior, 
which is a consequence of the large shell gap, indicates a first-order phase transition~\cite{luciano2014}.

A combinatorial quasi-particle model in the microcanonical ensemble describes surprisingly well the two level densities.
The adding of more quasi-particles by breaking of $J=0$ nucleon pairs is found to be the main mechanism for creating additional levels. 
According to the $\mu$CM, at $E=12$~MeV the number of quasi-particles are five and six for $^{89,90}$Y, respectively. 
For $^{90}$Y, the extra neutron outside $^{89}$Y behaves like a spectator and is responsible for $\approx6$ times higher 
level density. Furthermore, it is shown that the temperature-dependent combinatorial model with the D1M Gogny force (TDCG) 
fails to reproduce the experimental level densities.

The few levels in $^{89}$Y below $4-6$~MeV of excitation energy are responsible for a strongly fluctuating parity distribution. 
However, at the neutron separation energy both nuclei seem to reveal equally many positive and negative parity states.

It is very interesting to note that the level densities of $^{89,90}$Y seem to exhibit the same constant slope in a log scale,
corresponding to a common temperature of $T\approx 1.0$~MeV. If this trend persists when adding more neutrons, it may give guidelines 
on how to extrapolate level densities to the more neutron-rich isotopes.

\acknowledgements

The authors wish to thank J.C.~M{\"{u}}ller, E.A.~Olsen, A.~Semchenkov and J.~Wikne at the 
Oslo Cyclotron Laboratory for providing excellent experimental conditions. 
This work was supported by the Research Council of Norway (NFR).

\vfill
\end{document}